\documentclass[cite,twocolumn,aps,prb,amsmath,floatfix]{revtex4}
\usepackage{graphicx}
\usepackage{hyperref}
\usepackage{braket}
\usepackage{bm}

\def\prl{Phys. Rev. Lett. }
\def\prb{Phys. Rev. B }
\def\rmp{Rev. Mod. Phys. }
\def\ie{{\it i.e. }}

\def\etal{{\it et~al.}}
\def\ec{\epsilon_c}

\def\bn{\bar{n}}
\def\bnf{\bar{n}_f}
\def\rhos{\rho_s}
\def\H{\mathcal{H}}
\def\bnabla{\bm{\nabla}}
\def\br{{\bf r}}
\def\bA{{\bf A}}

\def\bE{{\bf E}}
\def\bJ{{\bf J}}

\def\byh{{\bf \hat{y}}}
\def\bzh{{\bf \hat{z}}}

\def\ton{{T_{\rm onset}}}
\def\LSCO{La$_{2-x}$Sr$_x$CuO$_4$ }

\begin{document}

\title{The Role of the Core Energy in the Vortex Nernst Effect}

\author{Gideon Wachtel}
\author{Dror Orgad}
\affiliation{Racah Institute of Physics,
The Hebrew University, Jerusalem 91904, Israel}
\begin{abstract}
  We present an analytical study of diamagnetism and transport in a
  film with superconducting phase fluctuations, formulated in terms of
  vortex dynamics within the Debye-H\"uckle approximation. We find
  that the diamagnetic and Nernst signals decay strongly with
  temperature in a manner which is dictated by the vortex core energy.
  Using the theory to interpret Nernst measurements
  of underdoped \LSCO above the critical temperature regime we obtain
  a considerably better fit to the data than a fit based on Gaussian
  order-parameter fluctuations. Our results indicate that the core
  energy in this system scales roughly with the critical
  temperature and is significantly smaller than expected from BCS theory. 
  Furthermore, it is necessary to assume that the vortex mobility is much 
  larger than the Bardeen-Stephen value in order to reconcile 
  conductivity measurements with the same vortex picture. 
  Therefore, either the Nernst signal is not due to superconducting phase 
  fluctuations, or that vortices in underdoped \LSCO have highly 
  unconventional properties.
\end{abstract}
\pacs{74.25.Fg, 74.40.-n, 74.72.-h}
\maketitle

Over the past decade the Nernst effect has become a widely used tool
in the study of strongly correlated electronic systems.  The Nernst
signal $e_N=E_y/(-\partial_x T)$, defined by the ratio between a
measured electric field $E_y$ and a transverse applied temperature
gradient $\partial_xT$ in an electrically isolated system subjected to
an external magnetic field $H_z$, is typically very small in
nonmagnetic normal metals.  Conversely, a much stronger effect may
arise in the flux-flow regime of superconductors, owing to the
transverse electric fields induced by the motion of vortices down the
temperature gradient. Consequently, the observation of a large Nernst
signal in the pseudogap state of the
cuprates\cite{Huebener,Ongxu,OngPRB01,Onglongprb} has been taken as evidence
that these systems support vortex-like superconducting fluctuations
over a wide temperature range above their critical temperature, $T_c$.
However, others have attributed the large Nernst signal to the response 
of quasiparticles in a symmetry-broken state competing with superconductivity.
\cite{Taillefer-Nature-stripes,Taillefer-broken-symm-Nature,Nernst-stripes}

Despite its appealing nature, the vortex based picture has not been
previously justified by an analytical treatment. However, several
studies have calculated the Nernst signal arising from superconducting
order-parameter fluctuations. The contribution of BCS Gaussian
fluctuations to the thermoelectric response of the normal state near
$T_c$ was obtained in Refs. \onlinecite{Ussishkin1,Ussishkin2}. This result
was subsequently extended to a wider range of temperatures and
magnetic fields\cite{Michaeli1,Michaeli2,Galitski}, as well as to
scenarios beyond that of BCS fluctuations.\cite{TDGLsim,Tan,Levchenko}
Experimentally, good agreement with the Gaussian theory was found in
amorphous Nb$_{0.15}$Si$_{0.85}$ films\cite{Pourret} and in overdoped,
but not underdoped cuprates\cite{Ussishkin1} (see, however, Ref. 
\onlinecite{Taillefer-Nature-Phys12}). 

A different approach, more pertinent to the present study, was taken
by Podolsky \etal\cite{Podolsky}, who built upon the
premise\cite{phase-nature} that in underdoped cuprates,
superconductivity is destroyed at $T_c$ by strong phase fluctuations,
whereas pairing correlations survive up to a considerably higher scale
$T_p$. Ignoring superconducting amplitude fluctuations the authors
calculated the Nernst signal in a stochastic two-dimensional (2D) $XY$
model via numerical simulations and a high-temperature expansion. In
addition, they devised a simulation method to calculate the
thermoelectric response based on vortex dynamics.\cite{Raghu}

In this Letter we aim to bridge the aforementioned theoretical gap and
present an analytical study of diamagnetism and transport in an
extreme type-II superconducting film 
that is formulated directly in terms of vortices.  We focus on
temperatures above $T_c$ where there is a finite density, $n_f$, of
free, unbound vortices. 
Our approach, which treats the vortex
interactions within a Debye-H\"uckle approximation, is inspired by
Ambegaokar \etal \cite{Ambegaokar} who considered vortex dynamics in
the context of superfluid films. A similar route was taken in the
study of the resistive transition of superconducting films by Halperin
and Nelson.\cite{Halperin}

Our treatment identifies the vortex core energy $\epsilon_c$ as an
important energy scale which controls the strong temperature
dependence of the fluctuation signals.
Using the theory we are able to obtain a fit to the transverse thermoelectric
response of underdoped \LSCO (LSCO) which is superior to the one based
on Gaussian fluctuations.  The available data imply that both
$\epsilon_c$ and $T_c$ share a similar doping dependence, with 
$\epsilon_c\approx 4-5T_c$. Such values are significantly lower than 
the Fermi energy, which is the expected $\epsilon_c$ from BCS theory. 
Moreover, in order to reconcile the vortex picture with conductivity 
data, one needs to assume that the vortex mobility is much larger than 
the Bardeen-Stephen value.\cite{Bardeen-Stephen} Thus, unless the 
strong Nernst and diamagnetic signals in underdoped LSCO are of 
non-superconducting origin, it appears that the vortex core is 
unconventional and plays an important role in this system.  


\emph{Vortex Hamiltonian and dynamics}.  A 2D superconductor, at
temperatures well below $T_p$ where the order parameter amplitude is
frozen, can be described by an $XY$-type Hamiltonian density of a
phase field $\theta$ coupled via its charge, $(2e<0)$, to an
electromagnetic vector potential $\bA$, and a constant superfluid
density $\rhos$:
\begin{equation}
  \label{eq:Hpsi}
  \H=(1+\psi)\left[\frac{\rhos}{2} \left(\bnabla\theta
    -\frac{2e}{\hbar c}\bA\right)^2 +\sum_i\ec\delta(\br-\br_i)\right].
\end{equation}
We assume that only vortices contribute to the otherwise uniform
$\bnabla\theta$. A vortex $i$ of vorticity $n_i=\pm1$ at coordinates
$\br_i=(x_i,y_i)$ contributes
\begin{equation}
  \label{eq:vor1}
  \bnabla\theta_i(\br)=n_i\bzh\times\bnabla\ln\frac{|\br-\br_i|}{r_0}
  =n_i\frac{\bzh\times(\br-\br_i)}{|\br-\br_i|^2},
\end{equation}
where $r_0$ is the vortex core radius, and $\bzh$ is a unit vector
perpendicular to the plane.  The continuum model and vortex
configuration, Eqs. (\ref{eq:Hpsi},\ref{eq:vor1}), are valid at scales
longer than $r_0$. Thus, a region of radius $r_0$ around $\br_i$ is
implicitly removed from the first term in Eq. (\ref{eq:Hpsi}).  Its
energy is given by the vortex core energy\cite{KT}, $\ec$, which we
assume to be constant across the sample.
Following Luttinger\cite{Luttinger}, we have introduced a
``gravitational'' field $\psi(\br)$ in order to study the response of
the system to a temperature gradient.

For concreteness, we consider a superconducting strip of infinite
extent along the $y$ direction, and of finite width $L$ in the $x$
direction. When needed, a constant transverse temperature gradient is
applied via $\psi(\br)=\psi' x$, and a uniform electric field
$\bE=E_y\byh$ is applied along the strip.  Working in the extreme
type-II limit we assume the presence of a uniform perpendicular
magnetic field $B\bzh$, and choose the gauge $\bA = \bA_0 + \bA_E$,
where $\bA_0=Bx\byh$, and $\bE=-\partial_t\bA_E/c$.
By symmetry, the average (over vortices' positions) phase gradient
$\braket{\bnabla\theta}$ is directed along the strip and is
independent of the $y$ coordinate.


We approach the model given by Eq. (\ref{eq:Hpsi}) within a mean-field
Debye-H\"uckle approximation, in which correlations between vortices
are ignored. This is possible at temperatures higher than the
Beresinskii-Kosterlitz-Thouless (BKT) transition temperature
$T_{BKT}$, for length scales longer that the Debye-H\"uckle screening
length $r_s$, where vortex interactrions are screened by thermally
excited vortices. The effective description at such scales is still
given by Eq. (\ref{eq:Hpsi}), provided that $\rhos$ and $\ec$ assume
renormalized values, which include contributions from the superflow at
shorter distances.\cite{MinnhagenRMP} Consequently, these parameters
become temperature dependent. Dynamics is introduced into the model by
assuming that the probability $P_i(\br_i,t)$ to find the $i$th vortex
at position $\br_i$ and time $t$ obeys a mean field Fokker-Planck
equation.\cite{supp} The corresponding probability current density for
vortrex $i$ is given by \cite{FPgradT}
\begin{equation}
  \label{eq:Ji}
  \bJ^i(\br_i,t) = -\mu P_i(\br_i,t)\braket{\bnabla_i H}_i-\mu T\bnabla_i
  P_i(\br_i,t),
\end{equation}
where $H=\int\! d^2r\,\H$, $\mu$ is the vortex mobility, $T$ the
temperature (here, and throughout $k_B=1$), $\bnabla_i$ is the gradient with respect to $\br_i$, and
$\braket{\cdots}_i$ denotes an average over the position of all
vortices besides $\br_i$. Near equilibrium this reproduces the
mean-field Debye-H\"uckle theory, provided one ignores fluctuations by
taking $\braket{(\bnabla\theta)^2}\approx(\braket{\bnabla\theta})^2$.
The residual effect of fluctuations is accounted for by renormalizing
$\rhos$ and $\ec$.\cite{MinnhagenRMP}

For convenience we define the mean field
$u(x)\equiv\braket{\partial_y\theta}/2\pi$ and $a(x)\equiv A_y/\phi_0$
where $\phi_0 =\pi\hbar c/e$ is the flux quantum.  Using these
definitions we find\cite{supp} that the $x$ component of the
probability current density of vortex $i$ is given by
\begin{eqnarray}
  \label{eq:Jix}
  J^i_x(x) &=& \mu P_i(x) \Big[4\pi^2\rhos n_i(1+\psi)(u-a)
  -\ec\partial_x\psi\Big]
  \nonumber \\
   & &\!-\mu T\partial_{x}P_i(x).
\end{eqnarray}
Similarly, the average vorticity current density along $x$ is
\begin{eqnarray}
  J^v_x(x)
   &=& \sum_in_iJ^i_x(x)
  \nonumber \\
  & = & \Big. 4\pi^2\rhos\mu n_f(1+\psi)(u-a) \nonumber \\  & &
  -\mu\ec\partial_x\psi\partial_xu - \mu T\partial_x^2u,  
  \label{eq:Jvx}
\end{eqnarray}
where
$\partial_xu(x)=n(x)=\sum_in_iP_i(x)$ is the mean vorticity, whose
bulk value, as shown below, is set by $B$, and
$n_f(x)=\sum_iP_i(x)$ is the density of free vortices.
Within the equilibrium Debye-H\"uckle approximation\cite{supp} it is
possible to show that
\begin{equation}
  \label{eq:nf(x)}
  n_f\simeq \sqrt{4r_0^{-4}e^{-2\epsilon_c/ T}+n^2},
\end{equation}
which establishes a strong dependence of $n_f$ on $T$, for small
$B$. The average $y$ component of the electric current density
$\bJ^e=-c\langle\delta\H/\delta\bA\rangle$ is given by
\begin{eqnarray}
  \label{eq:Je}
  J^e_y =  \frac{4\pi^2\rhos c}{\phi_0}(1+\psi)(u-a).
\end{eqnarray}
Thus, the first term in Eq. (\ref{eq:Jix}) is just the vortex drift
in response to the Magnus force it experiences in an electric current
$J^e_y$. Note, that {\it all} free vortices, and not only those
responsible for the excess vorticity, contribute to the vorticity
current, Eq. (\ref{eq:Jvx}), via their response to the Magnus force.
As a result, the strong temperature dependence of $n_f$ is also
reflected in the transport coefficients.

\emph{Equilibrium magnetization}.
In equilibrium $\psi=0$, $E_y=0$, and we must have $J^v_x=0$. We
therefore need to find $u_0(x)$ which solves the following equation
\begin{equation}
  \label{eq:equil}
  4\pi^2\rhos n_f(u_0-\bn x)-T\partial_x^2u_0=0,
\end{equation}
with $\bn$ defined such that $a(x)=Bx/\phi_0=\bn x$.
We solve this equation, for small $B$, by choosing boundary conditions
in which the vorticity $n(x)=\partial_xu(x)$ vanishes at $x=0$ and
$x=L$. In terms of the Debye-H\"uckle screening length,
$r_s^{-2}=4\pi^2\rhos n_f/T$, we find
\begin{equation}
  \label{eq:u0sol}
  u_0(x)=\bn\left[x+r_s\frac{e^{-x/r_s}-e^{-(L-x)/r_s}}{1+e^{-L/r_s}}\right].
\end{equation}
The deviation of $u_0$ from $\bar{n}x$ near the edge leads, according
to Eq. (\ref{eq:Je}), to edge currents. Their integral gives rise to
an average magnetization density
\begin{equation}
  \label{eq:mag}
  M_z=\frac{1}{c\mathcal{A}}\int dy\int_0^L dx\,xJ^e_y\simeq
  -\frac{TB}{\phi_0^2n_f},
\end{equation}
where $\mathcal{A}$ is the area of the strip.  Here, and in the
following, we ignore corrections of order $\mathcal{O}(r_s/L)$.
Similar expressions to Eq. (\ref{eq:mag}) were obtained in several
previous studies.\cite{Halperin,Benfatto-magnetic,Oganesyan}


\emph{Electric conductivity}.  In order to study the linear response
of the system to a weak perturbing field $E_y(\omega)e^{-i\omega t}$
we need to obtain the dynamics of $u(x,t)$.  By employing
translational invariance in the $y$ direction\cite{supp} one can show
that
\begin{equation}
  \label{eq:dudt}
  \frac{\partial u}{\partial t}=-J^v_x.
\end{equation}
This is a local version of the equation used in
Refs. \onlinecite{Ambegaokar,Halperin}.
Solving it using Eq. (\ref{eq:Jvx}), we find in the bulk
$u(x,t)=\bn x+u(\omega)e^{-i\omega t}$ where
\begin{equation}
  \label{eq:uE}
  u(\omega)=\frac{1}{1-i\omega\tau}\,\frac{cE_y(\omega)}{i\omega\phi_0},
\end{equation}
and where we have introduced the relaxation time
$1/\tau=4\pi^2\rhos\mu n_f$.
Eq. (\ref{eq:Je}) then implies an electric conductivity
\begin{equation}
  \label{eq:sigma}
  \sigma_s(\omega)=\frac{4e^2}{h}\frac{1}{h\mu n_f}
  \,\frac{1}{1-i\omega\tau}.
\end{equation}
This result is identical to the conductivity obtained by Halperin and
Nelson\cite{Halperin} for temperatures above $T_c$.


\emph{Thermoelectric coefficients}.  For systems with particle-hole
symmetry or when superconducting fluctuations dominate, the Nernst
signal is given by $e_N=\rho\alpha_{xy}=-\rho\alpha_{yx}$, where
$\alpha_{yx}$ is defined by
$J^e_y=\alpha_{yx}(-\partial_xT)$.\cite{Onglongprb}  Luttinger has
shown\cite{Luttinger} that $\alpha_{yx}$ can be deduced from the
response to a ``gravitational'' field $\psi$ according to the relation
$J^e_y=T\alpha_{yx}(-\partial_x\psi)$. Thus, we solve
Eq. (\ref{eq:dudt}) in the presence of
$\psi(x,t)=\psi'(\omega)xe^{-i\omega t}$. By writing
$u(x,t)=u_0(x)+u(\omega)e^{-i\omega t}$, where $u_0(x)$ is the
equilibrium solution of Eq. (\ref{eq:equil}), we find that to first
order in $\psi'(\omega)$
\begin{equation}
  \label{eq:upsi}
  \bar{u}(\omega)=\frac{1}{L}\int_0^Ldx\,u(x,\omega) =
  \frac{-M_z\phi_0n_f+\ec\bn}{1-i\omega\tau}\frac{\psi'(\omega)}{4\pi^2\rhos n_f}.
\end{equation}
Eq. (\ref{eq:Je}) leads then to the average electric current density
\begin{eqnarray}
  \overline{J^e_y}(\omega) & = & \frac{1}{\mathcal{A}}
  \int dy \int_0^L dx\,J^e_y(x,\omega) \nonumber \\
  & \simeq & \frac{-M_z\phi_0n_f+\ec\bn}{1-i\omega\tau}
  \frac{c\psi'(\omega)}{n_f\phi_0}+cM_z\psi'(\omega).
\end{eqnarray}
The response of $u(x,\omega)$ is given by the first term above. An
additional contribution, of opposite sign, comes from magnetization
currents near the edges.  Contrary to some previous
studies\cite{Ussishkin1,Podolsky} where this additional contribution
had to be subtracted\cite{Cooper}, in our treatment its opposite
effect is explicitly included in the second term.  In the DC limit,
$\omega\to 0$, we therefore obtain
\begin{eqnarray}
  \label{eq:alpha}
  \alpha_{yx}&=&-\frac{2ek_B}{h}\frac{B}{n_f\phi_0}
  \frac{\ec}{k_B T}
  = \frac{\ec}{T}\frac{cM_z}{T}.
\end{eqnarray}
This result should be compared with the constant ratio between
$\alpha_{yx}$ and $cM_z/T$, which was found for high temperatures in
Refs. \onlinecite{Ussishkin1,Podolsky} and \onlinecite{Raghu}.

Next, we consider the linear response ratio $\tilde\alpha_{xy}$
between an applied electric field and a transverse heat current
density, $J^Q_x=\tilde\alpha_{xy}E_y$.  We deduce $\bJ^Q$, which in
our model equals the energy current density, from the conservation
equation $\partial_t\mathcal{H}+\bnabla\cdot\bJ^Q=\bJ^e\cdot\bE$.
Its source term originates from the explicit time dependence of $\H$
via $\bA$. The result
\begin{equation}
  \label{eq:JE}
  \bJ^{Q} = -\rhos\left\langle\frac{\partial\theta}{\partial t}
  \left(\bnabla\theta-\frac{2e}{\hbar c}\bA\right)\right\rangle
  + \sum_i\ec\,\bJ^i,
\end{equation}
is consistent with the form used by Ussishkin \etal\cite{Ussishkin1},
once modified to include the energy current associated with the vortex
cores. If we additionally assume that the long superconducting strip
is periodic in the $y$ direction, then the $x$ component of the first
term in Eq. (\ref{eq:JE}) must vanish by symmetry, and we find that
Onsager's relation $\tilde\alpha_{xy}(B)=T\alpha_{yx}(-B)$ is obeyed.

\emph{Discussion}. Often (see Refs. \onlinecite{Huebener,Onglongprb} and
references therein), a phenomenological quantity called the vortex
transport entropy, $s_\phi$, is invoked in order to relate the
temperature gradient to the thermal force acting on a vortex,
\ie ${\bf f}=-s_\phi\bnabla T$. Based on Eq. (\ref{eq:Jix}) and
Luttinger\cite{Luttinger}, we identify $s_\phi=\ec/T$. For low
temperatures where there are no thermally excited vortices and the
flux-flow resistivity is the dominant form of damping, one can show by
neglecting vortex interactions\cite{Onglongprb} that
$\alpha_{yx}=-cs_\phi/\phi_0$. When taken together with the above
identification of $s_\phi$, this result is consistent with
Eq. (\ref{eq:alpha}), since at low temperatures $\bnf\phi_0=B$.

\begin{figure}[t!!!]
\centering
\includegraphics[width=\linewidth]{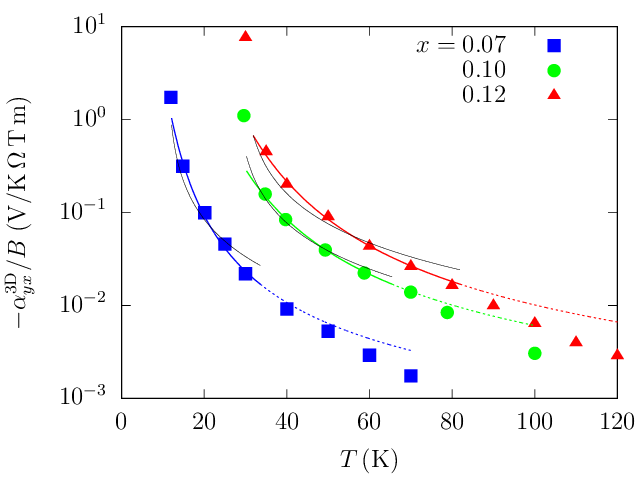}
\caption{ $-\lim_{B\rightarrow 0}\alpha^{\rm 3D}_{yx}/B=(\nu-\nu_n)$
of underdoped $\rm La_{2-{\it x}}Sr_{\it x}CuO_4$, where $\nu$ is the
Nernst coefficient, $\nu_n$ a subtracted background due to quasiparticles,
and $\rho$ is the in-plane resistivity.  The data for $x=0.07,0.10$
was extracted from Refs. \onlinecite{Ongxu,OngPRB01,Ando2004}, and for
$x=0.12$ from Ref. \onlinecite{Podolsky}.  The data was fitted to
Eqs. (\ref{eq:alpha}) (solid color curves).  In the regime indicated by
the dashed curves $r_0^2n_f>0.35$, and the theory is not expected to
be applicable.  The solid black curves depict the best fit to the
Gaussian fluctuations theory.\cite{Michaeli2,Galitski}}
\label{fig:al}
\end{figure}

As the temperature is raised through $T_{BKT}$, the density of free
vortices, $n_f$, rapidly increases. Our results,
Eqs. (\ref{eq:nf(x)},\ref{eq:mag},\ref{eq:alpha}), indicate that both
$M_z$ and $\alpha_{yx}$ should exhibit a consequent strong reduction
with temperature, much faster than the $1/T\ln(T/T_c)$ decay expected
from Gaussian fluctuations.\cite{Ussishkin1,Michaeli2,Galitski} To
look for such behavior in the cuprates we compare Eq. (\ref{eq:alpha})
divided by the LSCO layer separation, $d=6.5${\AA},
$\alpha^{\rm 3D}_{yx}=\alpha_{yx}/d$ with underdoped LSCO data.
According to Eq. (\ref{eq:nf(x)}), $n_f$ is determined by the renormalized
vortex core energy $\ec$, which reflects fluctuations at distances below
$r_s$ and is temperature dependent. For weak magnetic fields and
in the critical regime above $T_{BKT}$ this renormalization leads to
$n_F\sim\exp(-b/\sqrt{T-T_{BKT}})$\cite{MinnhagenRMP}, while at high temperatures
$n_F\sim\exp[-\ec/(T-\tilde b)]$.\cite{MinnhagenRMP,Benfatto-magnetic}
Here $b$ and $\tilde b$ are constants and $\ec$ is the bare core energy.
The lack of detailed knowledge about the the full temperature dependence
of $\ec$ allows for considerable freedom in the fitting procedure.
In order to constrain the fit, and since we are only interested in
a rough estimate of $\ec$, we choose to consider a constant $\ec$ and
also set $\phi_0/2\pi r_0^2=50\mbox{T}$.\cite{Podolsky}
Furthermore, we concentrate on the limit
$B\rightarrow 0$ and temperatures sufficiently above $T_c$, where
the renormalization effects are expected to be small, but low enough
so that vortices are distinct objects, \ie $r_0^2 n_f\ll 1$.
Figure \ref{fig:al} depicts the
measured $B\rightarrow 0$ limit of $-\alpha^{\rm 3D}_{yx}/B$ for LSCO
samples with $x\,(T_c)=0.07\,(11\mbox{ K})\,,0.10\,(27.5\mbox{ K})$
and $0.12\,(29\mbox{ K})$.  The solid color lines are the theoretical
fits in the temperature window $1.1T_c<T\lesssim 3T_c$, with a
constant $\ec$ as the only free fitting parameter. From these curves
we find $\ec\approx 58,\,114,\,143\,{\rm K}$, for the different doping
levels.  Comparable, but somewhat larger values, $\ec\approx 8T_c$,
were found by analyzing penetration depth measurements in underdoped
$\rm Y_{1-{\it x}}Ca_{\it x}Ba_2Cu_3O_{7-\delta}$ bilayer
films.\cite{Benfatto-Bilayer} For comparison we also include the best
fit to the data based on the theory of Gaussian
fluctuations.\cite{Michaeli2,Galitski} Clearly, the data exhibits a
faster decay than the Gaussian theory above the critical region around
$T_c$. In addition, we fitted the data to the high-$T$ result
$\alpha_{yx}\propto T^{-4}$ of the stochastic $XY$
model.\cite{Podolsky} We obtained a good fit for $x=0.12$, but found
overestimation of the data in the range $1.1T_c<T<2T_c$ $(3T_c)$ for
$x=0.10$ $(0.07)$.

The Nernst effect onset temperature, $\ton$, is defined as the
temperature for which the Nernst coefficient $\nu=e_N/B$ goes below a
threshold value, typically around $\nu=4\,{\rm nV/K\, T}$. Such levels
can be reached using Eq. (\ref{eq:alpha}) only if one takes
$r_0^2n_f\sim 1$. This, however, is beyond the validity of our theory.
Indeed, we find that the experimental data begin to deviate from the
theoretical curves at temperatures where $r_0^2n_f>0.35$, indicated by
dashed lines in Fig. \ref{fig:al}.  Thus, although our theory agrees
with the Nernst measurements up to $T\approx 3\,T_c$, it cannot
account for $\ton$, which is probably controlled by a combination of
lattice effects\cite{Podolsky} and amplitude
fluctuations.\cite{Ussishkin1}

The Nernst signal in the cuprate pseudogap regime exhibits a maximum
as a function of the magnetic field, which shifts to higher fields
with increasing
temperature.\cite{Onglongprb,Taillefer-Nature-Phys12} While we do not
have a theory for the maximum we note that
Eqs. (\ref{eq:nf(x)},\ref{eq:alpha}) imply a crossover, set by the
condition
$B/\phi_0\sim n_f(T,B=0)$, from a linear-$B$ dependence of
$\alpha_{yx}$ at weak fields towards saturation at higher
fields. Across this scale magnetic field-induced vortices dominate,
screening is reduced and correlation effects are enhanced, leading
potentially to the suppression of $\alpha_{yx}$.

In conclusion, we showed that within the vortex picture of phase
fluctuating superconductors, $\ec$ plays an essential role in the
thermoelectric response. The vortex core energy was also found to be
important in determining $T_c$ of layered
superconductors.\cite{Benfatto3d} Uncovering the role played by $\ec$
in other phenomena may help in identifying the physics underlying the
different temperature scales observed in the cuprates. Equally
pertinent is gaining an understanding of the factors which determine
$\ec$ itself. Here we briefly mention the need for a model of ``cheap
vortices'', in which vortices support a state close in energy to the
superconducting phase.\cite{Lee} It seems to us that the checkerboard
state observed around vortex cores\cite{Hoffman} is a natural
candidate.

Nevertheless, if the Nernst signal in underdoped cuprates is, in fact, 
due to thermally excited vortices, one must also understand why experiments 
do not show signatures of fluctuation enhanced conductivity over a similar 
temperature range. More specifically, if the vortex mobility is given by the 
Bardeen-Stephen result\cite{Bardeen-Stephen}, $\mu\approx 8\pi e^2 r_0^2 /h^2\sigma_n$, 
then Eq. (\ref{eq:sigma}) gives a fluctuation contribution $\sigma_s=\sigma_n/2\pi r_0^2 n_f$, 
where $\sigma_n$ is the normal state conductivity. This would imply, using our estimate 
$\epsilon_c\approx 4-5T_c$, from fitting the LSCO Nernst data, and 
Eq. (\ref{eq:nf(x)}), that $\sigma_s>\sigma_n$ for $T<2T_c$, in contradiction 
to experiments. To avoid such a contradiction within our model, we must therefore 
assume that $\mu$ is much larger than the Bardeen-Stephen value, thereby reducing $\sigma_s$ while 
not affecting $M_z$ and $\alpha_{yx}$. A similar conclusion regarding $\mu$ was reached based on 
THz time-domain spectroscopy in LSCO.\cite{fast-vortices} The above discussion further indicates that 
understanding the vortex core in the cuprates may call for physics beyond standard BCS theory.  


We would like to thank Daniel Podolsky for helpful discussions.  This
research was supported by the Israel Science Foundation (Grant
No. 585/13).

\begin{widetext}
  \newpage
  \begin{center}
    \large{\bf Supplemental material}
  \end{center}

\section{Debye-H\"uckle Approximation in Equilibrium}

At high temperatures, it is possible to study the vortex Hamiltonian
within the Debye-H\"uckle approximation, which is best formulated
using a variational mean-field approach. Assume that the state of the
system is defined by the vorticity at each lattice site, 
$n_\br=0,-1,+1$. In the variational mean-field ansatz the density
matrix is factored into a product of local probabilities,
\begin{equation}
  \rho=\prod_\br\rho_\br(n_\br),
\end{equation}
with the effect that the entropy is given by
\begin{equation}
  S = -{\rm Tr}\rho\ln\rho=-\sum_\br\sum_{n_\br}\rho_\br(n_\br)\ln
  \rho_\br(n_\br).
\end{equation}
Additionally, one approximate the average Hamiltonian by
\begin{equation}
  \label{eq:Hmf}
  \braket{H} \approx \frac{1}{2}\rhos\int d^2r(1+\psi)
  \left(\braket{\bnabla\theta}-\frac{2e}{\hbar c}\bA\right)^2
  +\ec\sum_\br(1+\psi(\br))\braket{|n_\br|},
\end{equation}\
while ignoring the contribution coming from fluctuations in $\bnabla\theta$,
\begin{equation}
  \braket{H_{\rm fluc.}} =\frac{1}{2}\rho_s\int d^2r(1+\psi)\left(\braket{
      (\bnabla\theta)^2}
    -\braket{\bnabla\theta}^2\right).
\end{equation}
$\braket{\bnabla\theta}$ is given by
\begin{equation}
  \braket{\bnabla\theta(\br)}=\overline{\bnabla\theta}+\sum_{\br'}
  \braket{n_{\br'}}\frac{\bzh\times(\br-\br')}{(\br-\br')^2},
\end{equation}
where $\overline{\bnabla\theta}$ is the uniform part of $\bnabla\theta(\br)$, 
which does not rise from vortices, 
\begin{equation}
  \braket{n_\br}=\sum_{n_\br}\rho_\br(n_\br)n_\br,
\end{equation}
and
\begin{equation}
  \braket{|n_\br|}=\sum_{n_\br}\rho_\br(n_\br)|n_\br|.
\end{equation}
$\rho_\br(n_\br)$ itself is determined by
minimizing the free energy $F=\braket{H}-TS$, with the constraint
\begin{equation}
  \sum_{n_\br}\rho_\br(n_\br)=1,
\end{equation}
\begin{equation}
  \frac{\partial F}{\partial\rho_\br(n_\br)}   =  \rhos\int d^2r'\left(
    \braket{\bnabla'\theta(\br')}-\frac{2e}{\hbar c}\bA(\br')\right)\cdot
  \frac{\bzh\times(\br'-\br)}{(\br'-\br)^2}n_\br 
  +\ec(1+\psi(\br)) |n_\br|  +T\ln\rho_\br(n_\br)=\alpha.
\end{equation}
Solving for $\rho_\br$ we find
\begin{equation}
  \rho_\br(n_\br) = \frac{1}{z_\br}e^{-\beta\ec|n_\br|-\beta\varphi(\br)n_\br},
\end{equation}
where
\begin{equation}
  \varphi(\br) = \rhos\int d^2r'\left(\braket{\bnabla'\theta(\br')}
    -\frac{2e}{\hbar c}\bA(\br')\right)\cdot\frac{\bzh\times(\br'-\br)}
  {(\br'-\br)^2},
\end{equation}
and
\begin{equation}
  z_\br=1+e^{-\beta\ec}2\cosh\beta\varphi(\br).
\end{equation}
For small $e^{-\beta\ec}$ we find
\begin{equation}
  \braket{|n_\br|} \approx e^{-\beta\ec}2\cosh\beta\varphi(\br),
\end{equation}
and
\begin{equation}
  \braket{n_\br} \approx -e^{-\beta\ec}2\sinh\beta\varphi(\br).
\end{equation}
Eliminating $\varphi$ gives
\begin{equation}
  \braket{|n_\br|}=\sqrt{4e^{-2\beta\ec}+\braket{n_\br}^2},
\end{equation}
which, after dividing through by $r_0^2$, reads
\begin{equation}
  n_f=\sqrt{4r_0^{-4}e^{-2\beta\ec}+n^2}.
\end{equation}

\section{Vortex Dynamics}

\subsection{Mean-Field Fokker-Planck equations}

In order to formulate dynamics of the vortices in our model, we assume
that the number of vortices is the same as in equilibrium, and that
their vorticity is fixed. Events of vortex-anti-vortex creation and
annihilation are important for non-linear response at $T_c$, but have
a negligible effect on linear response, and are therefore
ignored. Thus, it is possible to formulate vortex dynamics using a
Fokker-Planck equation for the positions of all vortices, $\{\br_i\}$,
each with a given vorticity $\{n_i=\pm 1\}$:
\begin{equation}
  \label{eq:F-P}
  \frac{\partial P(\{\br_i\},t)}{\partial t} = \sum_i\Big\{\mu\bnabla_i\cdot
    \left[ P(\{\br_i\},t)\bnabla_iH\right]+\mu T\nabla_i^2P(\{\br_i\},t)
  \Big\},
\end{equation}
where $\mu$ is the vortex mobility, $\bnabla_i$ is the gradient with respect
to $\br_i$, and $k_B=1$ is used throughout.
This is a complicated equation to solve, but it can be treated
approximately, in a manner similar to the Debye-H\"uckle approximation
in equilibrium, by factoring the probability density into a product
of single vortex probabilities,
\begin{equation}
  \label{eq:MF}
  P(\{\br_i\},t)=\prod_iP_i(\br_i,t).
\end{equation}
Integrating the left side of Eq. (\ref{eq:F-P}) over the positions of all
vortices aside from the position of the $i$th gives
\begin{eqnarray}
  \prod_{j\ne i}\int d^2r_j\frac{\partial P(\{\br_i\})}{\partial t} & = &
  \prod_{j\ne i}\int d^2r_j\sum_k\prod_{l\ne k}P_l(\br_l,t)\frac{\partial P_k(\br_k,t)}
  {\partial t} \nonumber \\ & = & P_i(\br_i,t)\sum_{k\ne i}\prod_{j\ne i,k}\left(\int d^2r_j
    P_j(\br_j,t)\right)\int d^2r_k\frac{\partial P_k(\br_k)}{\partial t} 
  + \frac{\partial P_i(\br_i,t)}{\partial t}\prod_{j\ne i}\left(\int d^2r_j
    P_j(\br_j,t)\right) \nonumber \\ & = & \frac{\partial P_i(\br_i,t)}{\partial t},
\end{eqnarray}
where we demand that the single vortex probabilities are normalized,
\begin{equation}
  \int d^2r_j P_j(\br_j,t) = 1.
\end{equation}
Preforming the same integral on the right side of the Fokker-Planck
equation  gives
\begin{eqnarray}
  \frac{\partial P_i(\br_i,t)}{\partial t} & = & \prod_{j\ne i}\int d^2r_j
  \mu\sum_k\bnabla_k\cdot
  \Big[P(\{\br_k\},t)\bnabla_kH(\{\br_k\})+T\bnabla_kP(\{\br_k\})\Big] \nonumber \\
  \nonumber & = & P_i(\br_i,t)\mu\sum_{k\ne i}\int d^2r_k\bnabla_k\cdot
  \Big[P_k(\br_k,t)\braket{\bnabla_kH}_{ik} +T\bnabla_kP_k(\br_k)\Big] 
  + \mu\bnabla_i\cdot\Big[P_i(\br_i,t)\braket{\bnabla_iH}_{i}
  +T\bnabla_iP_i(\br_i)\Big],\nonumber  \\
  \label{eq:FPintri}
\end{eqnarray}
where
\begin{equation}
  \braket{\bnabla_iH}_i=\prod_{j\ne i}\left(\int d^2r_jP_j(\br_j,t)\right)
  \bnabla_iH,
\end{equation}
and
\begin{equation}
  \braket{\bnabla_kH}_{ik}=\prod_{j\ne i,k}\left(\int d^2r_jP_j(\br_i,t)\right)
  \bnabla_kH.
\end{equation}
$\braket{\bnabla_kH}_{ik}$ is similar to $\braket{\bnabla_kH}_{k}$
except for an interaction term $H_{ik}$ between vortex $k$ and vortex~$i$:
\begin{equation}
  \label{eq:Hik}
  \braket{\bnabla_kH}_{ik} = \braket{\bnabla_kH}_{k}-\braket{\bnabla_kH_{ik}}_{k}
  + \bnabla_kH_{ik}
\end{equation}
Substituting Eq. \ref{eq:Hik} into Eq. \ref{eq:FPintri} we find that
the single vortex Fokker-Planck equation is
\begin{equation}
  \frac{\partial P_i(\br_i)}{\partial t}=\mu\bnabla_i\cdot\Big[P_i(\br_i,t)
  \braket{\bnabla_iH}_{i}   +T\bnabla_iP_i(\br_i)\Big],
\end{equation}
provided that
\begin{equation}
  \sum_{k\ne i}\int d^2r_k\bnabla_k\cdot\Big[P_k(\br_k,t)\bnabla_kH_{ik}
  -P_k(\br_k,t)\braket{\bnabla_kH_{ik}}_{k}\Big]=0.
\end{equation}
This can be shown to be the case on our strip where there is
translational invariance in the $y$ direction.

\subsection{Derivation of the vorticity current}

As shown above, the Fokker-Planck equation can be separated into
single vortex equations,
\begin{equation}
  \label{eq:F-Pi}
  \frac{\partial P_i(\br_i,t)}{\partial t} = \mu\bnabla_i\cdot\left[
    P_i(\br_i,t)\braket{\bnabla_iH}_i\right]+\mu T\nabla_i^2P_i(\br_i,t),
\end{equation}
where $\braket{-\bnabla_iH}_i$ is the force on vortex $i$, averaged
over the position of all other vortices
\begin{equation}
  \braket{\bnabla_iH}_i = \prod_{j\ne i}\left(\int d^2r_j\,P_j(\br_j,t)
    \right)\bnabla_iH(\{\br\}) = \bnabla_i\frac{\delta\braket{H}}{\delta P_i(\br_i)}.
\end{equation}
Various average quantities can be calculated using the single vortex
probability density
\begin{equation}
  P_i(\br,t)=\braket{\delta(\br-\br_i(t))},
\end{equation}
and the probability current density
\begin{equation}
  \bJ_i(\br,t)=\braket{\delta(\br-\br_i(t))\dot{\br}_i(t)}.
\end{equation}
Interpreting the single vortex Fokker-Planck equation as a probability
conservation condition, it is evident that
\begin{equation}
  \label{eq:Ji}
  \bJ_i(\br_i,t) = -\mu P_i(\br_i,t)\braket{\bnabla_iH}_i-\mu T\bnabla_i
  P_i(\br_i,t).
\end{equation}
Translational invariance in the $y$ direction (along the strip)
requires that $P_i$ and $\bJ_i$ are independent of the $y$
coordinate. For example, the vorticity can be wrriten as
\begin{equation}
  \partial_xu(x,t)=n(x,t)=\sum_i\braket{n_i\delta(\br-\br_i(t))}
  = \sum_in_iP_i(x,t),
\end{equation}
the free vortex density is
\begin{equation}
  n_f(x,t)=\sum_i\braket{\delta(\br-\br_i(t))}  = \sum_iP_i(x,t),
\end{equation}
and the vorticity current is given by
\begin{equation}
  J^v_x(x,t) = \sum_i\braket{n_i\delta(\br-\br_i(t))\dot{x}_i}
  = \sum_in_iJ_{i,x}(x,t).
\end{equation}
Ignoring the same fluctuation term in
$\braket{H}$ as in Eq. \ref{eq:Hmf}, we find
\begin{eqnarray}
  \frac{\partial \braket{H}_i}{\partial x_i} & = &
  \frac{\partial}{\partial x_i}\frac{\delta\braket{H}}{\delta P_i(\br_i)}
  \nonumber \\  & \approx & n_i\rhos\frac{\partial}
  {\partial x_i}\int d^2r'[1+\psi(x')]\left( \braket{\bnabla\theta(\br')}
    -\frac{2e}{\hbar c}\bA(\br')\right) \cdot\frac{\bzh\times(\br'-\br_i)}
  {(\br'-\br_i)^2}+\ec\frac{\partial}{\partial x_i}\psi(\br_i)
   \nonumber \\ & = & n_i\rhos\frac{\partial}{\partial x_i}\int dx'\,2\pi
  [1+\psi(x')][u(x')
  -a(x')]\int dy'\frac{x'-x_i}{(x'-x_i)^2+(y'-y_i)^2}+\ec\frac{\partial}
  {\partial x_i}\psi(x_i) \nonumber \\ & = &
  n_i\rhos\frac{\partial}{\partial x_i}\int dx'\,2\pi[1+\psi(x')][u(x')-a(x')]
  \,\pi\, {\rm sign}(x'-x_i)+\ec\frac{\partial} {\partial x_i}
    \psi(x_i) \nonumber \\ & = & -n_i4\pi^2\rhos[1+\psi(x_i)]
  [u(x_i)-a(x_i)]
  +\ec\frac{\partial}{\partial x_i}\psi(x_i).
\end{eqnarray}
Therefore, the vorticity current density is
\begin{eqnarray}
  J^v_x(x,t) & = & \sum_in_iJ_{i,x}(x,t) \nonumber \\
  & = & \sum_in_i\left[-\mu P_i(x_i,t)\frac{\partial\braket{H}_i}
    {\partial x_i}-\mu T\frac{\partial P_i(x_i,t)}{\partial x_i}
  \right]_{x_i=x} \nonumber \\ & = & \sum_in_i\left[\mu P_i(x_i,t)n_i4\pi^2\rhos
    [1+\psi(x_i)][u(x_i)-a(x_i)]-\mu P_i(x_i,t)\ec\frac{\partial}{\partial x_i}
    \psi(x_i) 
    -\mu T\frac{\partial P_i(x_i,t)}{\partial x_i} \right]_{x_i=x}
   \nonumber \\ & = & \sum_i\Big[4\pi^2\rhos\mu P_i(x,t)[1+\psi(x)][u(x)-a(x)]
      -\mu\ec\partial_x\psi(x)n_iP_i(x,t) 
    -\mu Tn_i\partial_xP_i(x,t)\Big],
\end{eqnarray}
which finally gives
\begin{equation}
  J^v_x = 4\pi^2\rhos\mu n_f (1+\psi)(u-a) -\mu\ec\partial_x\psi
  \partial_xu  -\mu T\partial_x^2u.
\end{equation}

\section{Dynamic equation for $u$}

In order to study the linear response of the system to weak, time dependent,
perturbing fields $\bE$ and $\bnabla\psi$, we must obtain the dynamics
of the field $u(x,t)$.
\begin{eqnarray}
  \frac{\partial u}{\partial t} & = & \frac{1}{2\pi}\left\langle
    \sum_i \dot x_i \frac{\partial}{\partial x_i}\partial_y\theta
  \right\rangle  \nonumber \\
  \int dy\,\frac{\partial u}{\partial t} & = & \frac{1}{2\pi}\left\langle
    \sum_i \dot x_i \frac{\partial}{\partial x_i}\int dy\,\partial_y\theta
  \right\rangle \nonumber \\  & = &
  \frac{1}{2\pi}\left\langle
    \sum_i \dot x_i \frac{\partial}{\partial x_i}n_i\pi{\rm sign}(x-x_i)
  \right\rangle \nonumber \\  & = &
   -\left\langle
    \sum_i \dot x_i n_i\delta(x-x_i)
  \right\rangle = 
  -\int dy\,J^v_x.
\end{eqnarray}
By translational invariance in the $y$ direction we find
\begin{eqnarray}
  \label{eq:dudt}
  \frac{\partial u}{\partial t}=-J^v_x.
\end{eqnarray}

\end{widetext}

\begin{thebibliography}{999}

%

\bibitem{Huebener} For a review of earlier results see,
R.~P.~Huebener, Supercond. Sci. Technol. {\bf 8}, 189 (1995).

\bibitem{Ongxu} Z.~A.~Xu, N.P.~Ong, Y.~Wang, T.~Kakeshita, and S.~Uchida,
Nature (London) {\bf 406}, 486 (2000).

\bibitem{OngPRB01} Y.~Wang, Z.~A.~Xu, T.~Kakeshita, S.~Uchida, S.~Ono,
Y.~Ando, and N.~P.~Ong, \prb {\bf 64}, 224519 (2001).

\bibitem{Onglongprb} Y.~Wang, L.~Li, and N.~P.~Ong,
\prb {\bf 73}, 024510 (2006).

\bibitem{Taillefer-Nature-stripes} O.~Cyr-Choini${\rm \grave{e}}$re,
  R.~Daou, F.~Lalibert${\rm \grave{e}}$, D.~LeBoeuf,
  N.~Doiron-Leyraud, J.~Chang, J.-Q.~Yan, J.-G.~Cheng, J.-S.~Zhou,
  J.~B.~Goodenough, S.~Pyon, T.~Takayama, H.~Takagi, Y.~Tanaka, and
  L.~Taillefer, Nature (London) {\bf 458}, 743 (2009).

\bibitem{Taillefer-broken-symm-Nature} R.~Daou, J.~Chang, D.~LeBoeuf,
  O.~Cyr-Choini${\rm \grave{e}}$re, F.~Lalibert${\rm \grave{e}}$,
  N.~Doiron-Leyraud, B.~J.~Ramshaw, R.~Liang, D.~A.~Bonn, W.~N.~Hardy,
  and L.~Taillefer, Nature (London) {\bf 463}, 519 (2010).

\bibitem{Nernst-stripes} A.~Hackl, M.~Vojta, and S.~Sachdev, 
\prb {\bf 81}, 045102 (2010).

\bibitem{Ussishkin1} I.~Ussishkin, S.~L.~Sondhi and D.~A.~Huse,
\prl {\bf 89} 287001 (2002).

\bibitem{Ussishkin2} I.~Ussishkin, \prb {\bf 68}, 024517 (2003).

\bibitem{Michaeli1} K.~Michaeli and A.~M.~Finkel'stein, \prb {\bf 80},
214516 (2009).

\bibitem{Michaeli2} K.~Michaeli and A.~M.~Finkel'stein,
Europhys. Lett. {\bf 86}, 27007 (2009).

\bibitem{Galitski} M.~N.~Serbyn, M.~A.~Skvortsov, A.~A.~Varlamov,
and V.~Galitski, \prl {\bf 102}, 067001 (2009).

\bibitem{TDGLsim} S.~Mukerjee and D.~A.~Huse, \prb {\bf 70}, 014506 (2004).

\bibitem{Tan} S.~Tan and K.~Levin, \prb {\bf 69}, 064510 (2004).

\bibitem{Levchenko} A.~Levchenko, M.~R.~Norman, and A.~A.~Varlamov,
\prb {\bf 83}, 020506(R) (2011).

\bibitem{Pourret} A.~Pourret, H.~Aubin, J.~Lesueur, C.~A.~Marrache-Kikuchi,
L. Berg${\rm \acute{e}}$, L.~Dumoulin, and K.~Behnia, Nat. Phys. {\bf 2},
683 (2006).

\bibitem{Taillefer-Nature-Phys12} J.~Chang, N.~Doiron-Leyraud,
O.~Cyr-Choini${\rm \grave{e}}$re, G.~Grissonnanche,
F.~Lalibert${\rm \grave{e}}$, H.~Hassinger, J.~Ph.~Reid, R.~Daou,
S.~Pyon, T.~Takayama, H.~Takagi, and L.~Taillefer, Nat. Phys. {\bf 8},
751 (2012).

\bibitem{Podolsky} D.~Podolsky, S.~Raghu, and A.~Vishwanath,
\prl {\bf 99} 117004 (2007).

\bibitem{phase-nature} V.~J.~Emery and S.~A.~Kivelson,
Nature (London) {\bf 374}, 434 (1995).

\bibitem{Raghu} S.~Raghu, D.~Podolsky, A.~Vishwanath, and D.~A.~Huse,
\prb {\bf 78}, 184520 (2008).



\bibitem{Ambegaokar} V.~Ambegaokar, B.~I.~Halperin, D.~R.~Nelson, and
  E.~D.~Siggia, \prb {\bf 21} 1806 (1980).

\bibitem{Halperin} B.~I.~Halperin and D.~R.~Nelson, J. Low Temp. Phys.
  {\bf 36}, 599 (1979).

\bibitem{Bardeen-Stephen} J.~Bardeen and M.~J.~Stephen,
  Phys. Rev. {\bf 140}, A1197 (1965).

\bibitem{KT} J.~M.~Kosterlitz and D.~J.~Thouless, J. Phys. C {\bf 6},
1181 (1973).

\bibitem{Luttinger} J.~M.~Luttinger, Phys. Rev. {\bf 135}, A1505 (1964).

\bibitem{MinnhagenRMP} P.~Minnhagen, \rmp {\bf 59}, 1001 (1987).

\bibitem{supp} See online supplemental material for details.

\bibitem{FPgradT} The effect of a temperature gradient on a vortex
  enters this Fokker-Planck equation via the ``gravitational'' field
  $\psi$ in the Hamiltonian. Alternatively, it is possible to
  introduce a position dependent temperature, $T(\br)$, directly into
  the Fokker-Plank equation, in a manner which depends on the
  underlying microscopic dynamics. Assuming a vortex behaves like a
  Brownian particle, the corresponding probability current
  density\cite{vanKampen} is $\bJ_i(\br_i,t)=-\mu
  P_i(\br_i,t)\braket{\bnabla_iH}_i
  -\mu\bnabla\left[T(\br_i)P(\br_i,t)\right]$. This approach
  reproduces the same results we get by calculating the response to
  $\psi$, assuming that the density of free vortices is at local
  equilibrium, $n_f(\br)\sim\exp(-\ec/T(\br))$.

\bibitem{vanKampen} N.~G.~van~Kampen, J. Phys. Chem. Solids {\bf 49},
  673 (1988).

\bibitem{Benfatto-magnetic} L.~Benfatto, C.~Castellani, and
  T.~Giamarchi, \prl {\bf 99}, 207002 (2007).

\bibitem{Oganesyan} V.~Oganesyan, D.~A.~Huse, and S.~L.~Sondhi, \prb
  {\bf 73}, 094503 (2006).

\bibitem{Cooper} N.~R.~Cooper, B.~I.~Halperin, and I.~M.~Ruzin, \prb
  {\bf 55}, 2344 (1997).

\bibitem{Ando2004} Y.~Ando, S.~Komiya, K.~Segawa, S.~Ono, and
  Y.~Kurita, \prl {\bf 93}, 267001 (2004).

\bibitem{Benfatto-Bilayer} L.~Benfatto, C.~Castellani, and
  T.~Giamarchi, \prb {\bf 77}, 100506(R) (2008).

\bibitem{Benfatto3d} L. Benfatto, C. Castellani, and T. Giamarchi,
  \prl {\bf 98}, 117008 (2007).

\bibitem{Lee} P.~A.~Lee, N.~Nagaosa, and X.~Wen, \rmp {\bf 78}, 17
  (2006).

\bibitem{Hoffman} J.~E.~Hoffman, E.~W.~Hudson, K.~M.~Lang,
  V.~Madhavan, H.~Eisaki, S.~Uchida, and J.~C.~Davis, Science {\bf
    295}, 466 (2002).

\bibitem{fast-vortices} L.~S.~Bilbro, R.~Vald\'es Aguilar,
  G.~Logvenov, I.~Bozovic, and N.~P.~Armitage, \prb {\bf 84},
  100511(R) (2011).

\end{thebibliography}
\end{document}